\documentclass[conference]{IEEEtran}
\IEEEoverridecommandlockouts
\usepackage{setspace}
\usepackage{cite}
\usepackage{graphicx}
\usepackage{caption}
\usepackage{subcaption} 
\usepackage[cmex10]{amsmath}
\usepackage{mathtools}
\usepackage{amsthm}
\usepackage{amsfonts}
\usepackage[font=small,skip=0pt]{caption}
\usepackage{amssymb}
\usepackage{textcomp}
\usepackage{xcolor}
\def\BibTeX{{\rm B\kern-.05em{\sc i\kern-.025em b}\kern-.08em
    T\kern-.1667em\lower.7ex\hbox{E}\kern-.125emX}}
\usepackage{algorithm}
\usepackage{algpseudocode}
\usepackage{bm,xstring}
\usepackage{fixltx2e}
\usepackage{tikz}
\usetikzlibrary{shapes,arrows,positioning}
\usetikzlibrary{decorations.pathreplacing}

\DeclareMathOperator*{\argmax}{argmax}

\usepackage{color}

\newcommand{\ignore}[1]{}

\newtheorem{lemma}{Lemma}

\newtheorem{theorem}{Theorem}

\theoremstyle{definition}

\newtheorem{definition}{Definition}

\begin{document}

\title{Context-aware Status Updating: Wireless Scheduling for Maximizing Situational Awareness in Safety-critical Systems\\
}

\author{\IEEEauthorblockN{Tasmeen Zaman Ornee}
\IEEEauthorblockA{\textit{Dept. of ECE} \\
\textit{Auburn University}\\
Auburn, AL, USA \\
tzo0017@auburn.edu}
\and
\IEEEauthorblockN{Md Kamran Chowdhury Shisher}
\IEEEauthorblockA{\textit{Dept. of ECE} \\
\textit{Auburn University}\\
Auburn, AL, USA \\
mzs0153@auburn.edu}
\and
\IEEEauthorblockN{Clement Kam}
\IEEEauthorblockA{\textit{Information Technology Division} \\
\textit{U.S. Naval Research Laboratory}\\
Washington, DC, USA \\
clement.kam@nrl.navy.mil}
\and
\IEEEauthorblockN{Yin Sun}
\IEEEauthorblockA{\textit{Dept. of ECE} \\
\textit{Auburn University}\\
Auburn, AL, USA \\
yzs0078@auburn.edu}
\thanks{This work was supported in part by NSF grant CNS-2239677, ARO grant W911NF-21-1-0244, and ONR.}
}

\maketitle

\begin{abstract}
 In this study, we investigate a context-aware status updating system consisting of multiple sensor-estimator pairs. A centralized monitor pulls status updates from multiple sensors that are monitoring several safety-critical situations (e.g., carbon monoxide density in forest fire detection, machine safety in industrial automation, and road safety). Based on the received sensor updates, multiple estimators determine the current safety-critical situations. Due to transmission errors and limited communication resources, the sensor updates may not be timely, resulting in the possibility of misunderstanding the current situation. In particular, if a dangerous situation is misinterpreted as safe, the safety risk is high. In this paper, we introduce a novel framework that quantifies the penalty due to the unawareness of a potentially dangerous situation. This situation-unaware penalty function depends on two key factors: the Age of Information (AoI) and the observed signal value. For optimal estimators, we provide an information-theoretic bound of the penalty function that evaluates the fundamental performance limit of the system. 
To minimize the penalty, we study a pull-based multi-sensor, multi-channel transmission scheduling problem.
Our analysis reveals that for optimal estimators, it is always beneficial to keep the channels busy. Due to communication resource constraints, the scheduling problem can be modelled as a Restless Multi-armed Bandit (RMAB) problem. By utilizing relaxation and Lagrangian decomposition of the RMAB, we provide a low-complexity scheduling algorithm which is asymptotically optimal. Our results hold for both reliable and unreliable channels. Numerical evidence shows that our scheduling policy can achieve up to 100 times performance gain over periodic updating and up to 10 times over randomized policy. 

\end{abstract}


\section{Introduction}

A broad range of safety-critical systems is ubiquitous across the world. 
For instance, in industrial automation, it is essential to continuously monitor the safety of various machines \cite{grau2017industrial}. In patient health monitoring, precise tracking of the glucose level or the heart rate is imperative to swiftly implement precautionary measures when they are required \cite{abdulmalek2022iot}. In disaster monitoring, it is important to promptly monitor any consistent changes in temperature or humidity, as they could indicate a possible disaster  
\cite{seenivasan2015disaster}. In these safety-critical situations, the monitoring system needs timely access and accurately interpret the states of remote systems. Any misunderstanding of the system state can lead to severe consequences.

In practice, multiple sensors are required to track various safety-critical situations.
One challenge to continuously monitor these sensor measurements in real-time is the limited capacity of the communication medium. 
Moreover, some sensors may have more crucial content than others and hence need more attention. 
In this context, we adopt a pull-based system \cite{li2020waiting}  where a centralized monitor selects sensors and requests information when required. This selective retrieval of information ensures that the system receives essential information promptly while minimizing unnecessary resource consumption.

In this paper, we consider a discrete-time pull-based status updating system consisting of multiple sensors monitoring the status of different safety-critical situations. At every time slot, the selected sensors transmit their updates to a receiver through multiple unreliable channels. In the receiver, multiple estimators utilize the sensor updates to determine the current status of the safety-critical situations. Due to transmission errors, the sensor updates may not be fresh. One performance metric that characterizes data freshness is the age of information (AoI) 
\cite{KaulYatesGruteser-Infocom2012}. Let $U(t)$ be the generation time of the freshest observation delivered to the receiver by time $t$. The AoI, as a function of $t$, is defined as $\Delta(t) = t- U(t)$ which exhibits a linear growth with time $t$ and drops down to a smaller value whenever a fresher observation is delivered. In many real-time applications, it is important to consider AoI for making the scheduling decision. However, the time difference represented by AoI can only capture the timeliness of the information but it cannot capture its significance. This is particularly relevant in safety-critical situations where misunderstanding about the situation can lead to significant performance loss. Hence, relying solely on AoI-based decision-making is not perfect. If we consider signal observation along with AoI in decision-making, then the incurred performance loss can be significantly improved. 
One key observation in this study is that any misinterpretation of a dangerous situation yields a higher loss compared to the misinterpretation of a safe situation. Based on the above-mentioned insights, we introduce a framework for quantifying the \emph{cost of a dangerous situation} that characterizes the performance loss caused by situational unawareness.

The goal of this paper is to find the optimal scheduling policy to select sensors and to request observations while improving the system performance. The contributions of this paper are as follows:

\begin{itemize}
\item We introduce a novel framework for estimating the current status of a safety-critical system. In this framework, we adopt a general loss function $L(y, \hat y)$ that quantifies the incurred loss in wrongly estimating the actual safety level $y$ as $\hat y$. 
The loss $L$(\emph{dangerous, safe}) is higher than $L$(\emph{safe, dangerous}).
This distinction can not be captured by the traditional loss functions such as 0-1 loss, quadratic loss, and logarithmic loss. 
By adopting appropriate loss functions $L$, our framework can be applied to health, safety, and security monitoring. 

\item 
To measure the performance of safety-critical systems, we propose a penalty function that represents the expected loss $L$ given the AoI and the latest observation (see Section \ref{information_theory_view}). 
We also provide an information-theoretic lower bound of the penalty function by using $L$-conditional entropy \cite{dawid1998coherent, farnia2016minimax, Shisher2022}.   
This bound represents the fundamental performance limit of a safety-critical system. The entropy-based freshness metric in our study can significantly contribute to real-time applications such as estimation, inference, and perception. Earlier metrics such as Age of Incorrect Information (AoII) \cite{Ali_AoII}, Age of Synchronization (AoS) \cite{zhong2018two}, Urgency of Information (UoI) \cite{zheng2020urgency}, Version AoI \cite{yates2021age}, AoI at Query (QAoI) \cite{holm2021freshness}, Value of Information (VoI) \cite{VoI_Kosta}, and Uncertainty of Information (UoI) \cite{Chen_2022} did not quantify the fundamental performance limit of real-time applications. 
Moreover, most of the prior studies \cite{Ali_AoII,zhong2018two,zheng2020urgency,yates2021age,holm2021freshness} exhibit a monotonic relationship with AoI, whereas some recent studies show that the performance of real-time applications may degrade 
non-monotonically with AoI \cite{shisher2021age, Shisher2022, Chen_2022}. Our penalty function also allows the non-monotonic behavior with respect to AoI.

\item We consider a multi-sensor, multi-channel pull-based status updating problem. Our findings demonstrate that when utilizing one-time slot transmission time and optimal estimators, it is always beneficial to keep the channels busy (see Theorem \ref{theorem1}). However, channel resource limitations prevent all sensors from transmitting  information continuously. To address this issue, we formulate the multi-sensor, multi-channel transmission scheduling problem as a Restless Multi-armed Bandit (RMAB). We utilize relaxation and Lagrangian method to decompose the original problem into multiple separated Markov Decision Processes (MDPs). We solve each MDP by dynamic programming \cite{bertsekas2011dynamic}. By utilizing the solution to the MDPs, we provide
a low-complexity scheduling policy which is asymptotically optimal and the developed policy works for both reliable and unreliable channels. In \cite{chen2022index}, the authors proved asymptotic optimality under concave penalty functions and for optimal estimators. In contrast to \cite{chen2022index}, our result holds for for arbitrary estimators without any concavity condition.

\item Numerical results illustrate that our multi-sensor, multi-channel scheduling policy achieves up to 100 times performance gain over periodic updating policy and up to 10 times over randomized policy which randomly select sensors depending on the number of available channels.
 
\end{itemize}
\section{Related Work}

There exists a large number of studies on minimizing linear and nonlinear AoI functions \cite{SunNonlinear2019, Sun_TIT_2017, Bedewy_2021, Ornee2021, Wiener_TIT, jsac_survey}. One limitation of AoI is that it only captures the timeliness of the information while neglecting the actual influence of the conveyed information. In order to address this, several performance metrics were introduced in conjunction with AoI \cite{zhong2018two, Ali_AoII, zheng2020urgency, yates2021age, holm2021freshness, pappas2021goal, Chen_2022}. Age of Incorrect Information (AoII) was introduced in \cite{Ali_AoII} that is represented by a function of the age and the estimation error. In \cite{zhong2018two}, Age of Synchronization (AoS) was considered along with AoI to measure the freshness of a local cache. In \cite{zheng2020urgency}, the authors proposed Urgency of Information (UoI) that captures the context-dependence of the status information along with AoI. Version AoI was introduced in \cite{yates2021age} that represents how many versions are out-of-date at the receiver, compared to the transmitter. An AoI at Query (QAoI) metric was introduced in \cite{holm2021freshness} to capture the freshness only when required in a pull-based communication system. 
In addition, several research papers studied information-theoretic measures to evaluate the impact of information freshness along with information content \cite{VoI_Kosta, SunNonlinear2019, wang2022framework, soleymani2016optimal, Chen_2022}. In \cite{VoI_Kosta, SunNonlinear2019, wang2022framework, soleymani2016optimal}, the authors employed Shannon's mutual information to quantify the information carried by received data messages regarding the current signal value at the source and used Shannon’s conditional entropy to measure the uncertainty about the current signal value. Based on the studies of \cite{VoI_Kosta, SunNonlinear2019, wang2022framework, soleymani2016optimal}, the authors in \cite{Chen_2022} termed Uncertainty of Information (UoI) by using the Shannon's entropy. However, there exists a disparity between these information-theoretic metrics and the performance of real-time applications such as remote estimation and inference. In \cite{shisher2021age, Shisher2022, shisher2023learning}, a generalized conditional entropy associated with a loss function $L$, or $L$-conditional entropy $H_L(Y_t|X_{t-\Delta(t)})$ was utilized to address this disparity, where $Y_t$ is the true state of the source and $X_{t-\Delta(t)}$ is the observed value. Building upon the insights of \cite{shisher2021age, Shisher2022, shisher2023learning}, we utilized $L$-conditional entropy $H_L(Y_t|X_{t-\Delta(t)}=x, \Delta(t)=\delta)$ given both the AoI $\delta$ and the observed value $x$ to measure the impact of the AoI and the information content in remote estimation and inference.


In addition, there exists numerous papers on AoI-based sampling and scheduling \cite{SunNonlinear2019, kadota2018scheduling, Bedewy_2021, hsu2018age, Shisher2022, ornee2023whittle, xiong2022index, zou2021minimizing, chen2021scheduling, shisher2023learning}. 
In \cite{SunNonlinear2019}, sampling policies for optimizing non-linear AoI functions were studied. 
A joint sampling and scheduling problem to minimize monotonic AoI functions was considered in \cite{Bedewy_2021}. A Whittle index-based scheduling algorithm to minimize AoI for stochastic arrivals was considered in \cite{hsu2018age}. In \cite{Chen_2022}, the authors proposed a Whittle index-based scheduling policy to minimize the UoI modeled as Shanon entropy. Optimal scheduling policies for both single and multi-source systems were studied and a Whittle index policy was proposed for multi-source cases in \cite{Shisher2022}. A Whittle index policy for both signal-aware and signal-agnostic scheduling was reported in \cite{ornee2023whittle}. A remote estimation system with multiple IoT sensors monitoring multiple Wiener processes was studied and a Max-$k$ policy was proposed in \cite{yun2023remote}. 
Besides Whittle index-based policies that require an indexability condition, non-indexable scheduling policies were also studied in \cite{xiong2022index, zou2021minimizing, chen2021scheduling, chen2022index, shisher2023learning}. In this paper, because of the complicated nature of state transition along with erasure channels, we do not provide indexability. However, we provide a “Net-gain Maximization Policy” developed in \cite{shisher2023learning, chen2022index}. In addition, by utilizing information-theoretic approach, we show that it is always beneficial to keep the channels busy for optimal estimators.
Our scheduling policy is designed for pull-based communication model where the scheduling decision is based on AoI and observed signal and the developed policy is asymptotically optimal.
\section{Model, Metric, and Formulation}\label{model}

\begin{figure}
\centering
\includegraphics[width=9cm]{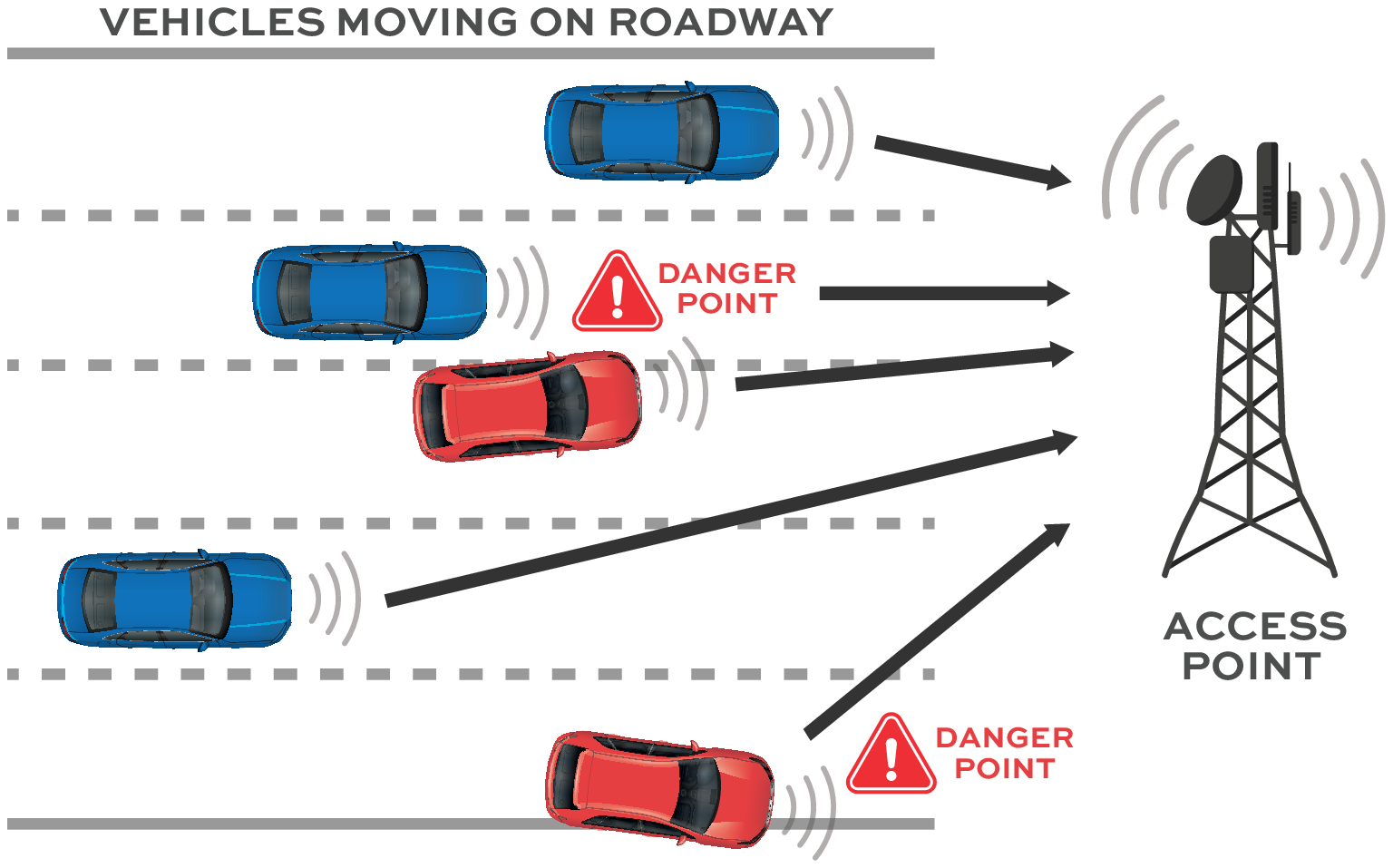}  
\caption{A multi-sensor, multi-channel vehicle safety monitoring system.}\vspace{-0.0cm}
\label{fig_model1}
\vspace{-0.3cm}
\end{figure}

\subsection{System Model}
Let us consider the status updating system depicted in Figure \ref{fig_model1}, where $N$ sensors transmit crucial status updates through $M$ unreliable wireless channels to a common receiver. Each sensor $n$ monitors a Markov signal $X_{n,t}$ representing the status of a safety-critical system. For instance, $X_{n,t}$ might represent the position of a vehicle on the road, or the joint angles of a robotic arm within a factory environment. A hazardous situation arises when the vehicle veers off the road or the robotic arm approaches a nearby object. We use $Y_{n,t}$ to quantify the level of danger for the safety-critical system, which is a function of the system status $X_{n,t}$. In practice,  $Y_{n,t}$ can be used to represent whether the vehicle is encroaching upon the road shoulder or the spatial distance between the robotic arm and the object. The receiver estimates the danger level signals $Y_{n,t}$ to ensure awareness of the hazards in the safety-critical systems. 

We consider a pull-based updating mechanism where the receiver requests status updates from the sensors whenever it is unsure about the situation. In response to the pull request, each sensor $n$ generates and submits a time-stamped updating message $(X_{n,t},t)$ to one wireless channel. We assume that it takes one-time slot for the transmission of a message update to the receiver. Due to wireless channel fading, the transmission of the status updates becomes unreliable. Let $p_n$ be the probability of a successful transmission from sensor $n$, irrespective of the selected wireless channel.



Due to transmission errors, the information received by the receiver will be stale and is represented by $X_{n, t-\Delta_n (t)}$ that is generated $\Delta_n (t)$ times ago.
The time different $\Delta_n (t)$ is usually called \emph{age of information (AoI)} \cite{KaulYatesGruteser-Infocom2012}, which represents the staleness of the status of the $n$ safety-critical system available at the receiver. 
At each time slot $t$, the AoI evolution of the $n$-th system is given by
\begin{align}
\Delta_n (t+1)=
\begin{cases}
\Delta_n (t) + 1, & \text{with probability $1-p_{n}$},\\
1, & \text{with probability $p_{n}$}.
\end{cases}
\end{align}

\begin{figure*}[ht]
  \centering
  \begin{subfigure}[t]{0.45\textwidth}
\includegraphics[width=\textwidth]{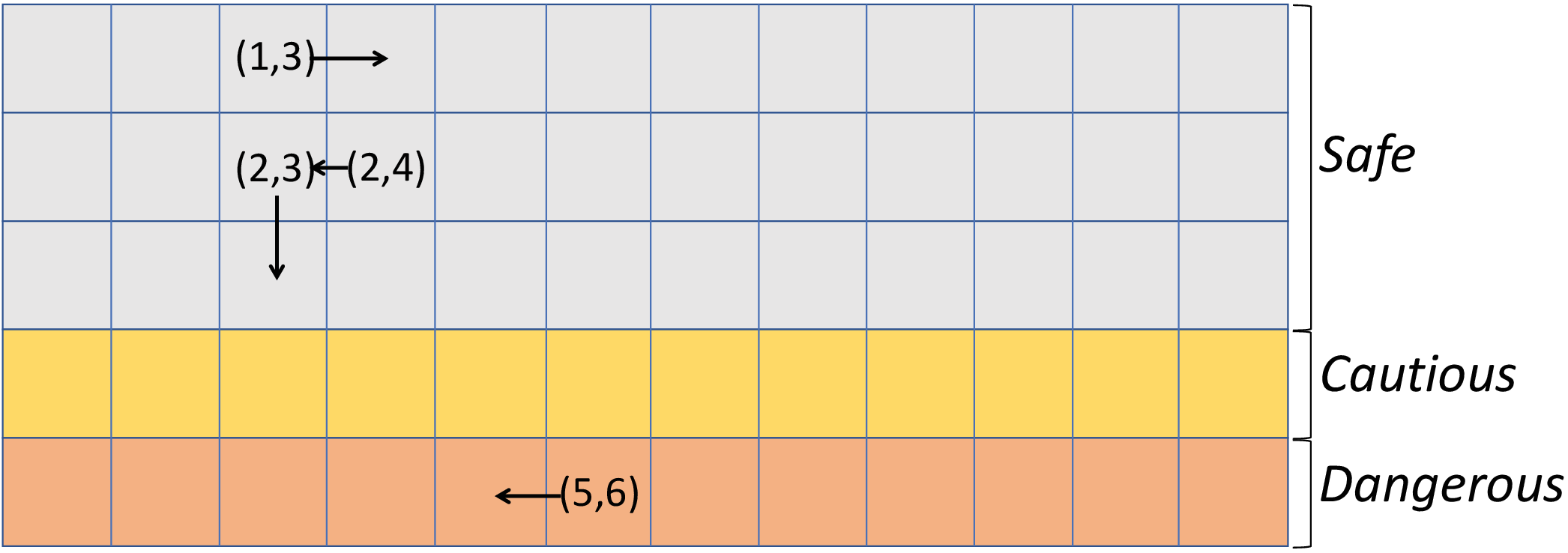}
  \subcaption{Gridworld}
\end{subfigure}
  \hspace{3mm}
\begin{subfigure}[t]{0.30\textwidth}
\includegraphics[width=\textwidth]{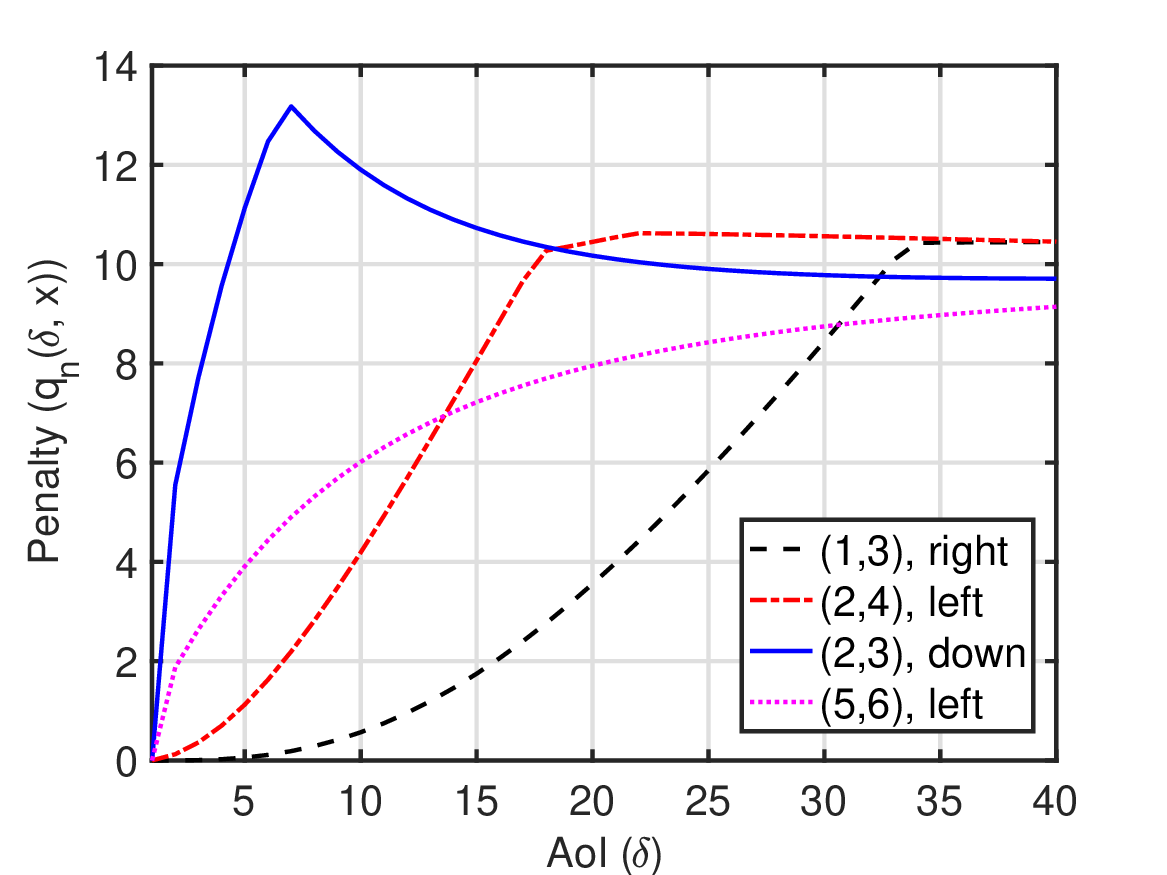}
  \subcaption{Penalty vs AoI}
\end{subfigure}
\hspace{4mm}
\vspace{4mm}
\caption{(a) Gridworld environment and (b) Penalty $(q_n (\delta, x))$ vs AoI ($\delta$) for four different given observation.}
\label{figure1}
\end{figure*}

\subsection{Loss Model for Situational Awareness}
Based on the latest available information, the $n$-th estimator outputs $a = \phi_n (\Delta_n (t), X_{n, t- \Delta_n (t)}) \in \mathcal{A}$, where $\phi : \mathbb {N} \times \mathcal{X} \to \mathcal A$ is a function of AoI $\Delta_n (t) \in \mathbb{N}$ and the received observation $X_{n, t- \Delta_n (t)} \in \mathcal X$.
%
The danger associated with the safety-critical system is characterized by a loss function $L : \mathcal Y \times \mathcal A \to \mathbb{R}$, where $L(y, a)$ is the incurred loss if $Y_{n,t} = y$ is the actual safety level of the surrounding environment and $a$ is estimated output of the safety level. Essentially, $L(\cdot, \cdot)$ serves as a metric to assess the \emph{cost of potential danger} within a safety-critical system.
To better understand the behavior of the loss $L$, we provide the following example that illustrate the impact of wrong estimation of the system state on $L$.

{\it \textbf{Example} Consider a road safety monitoring system that detects instances when any car veers off the road.
The sensors need to operate with high sensitivity to accurately communicate the measured variables in real-time. Let $Y_{n,t} =$ \emph\{danger, safe\} denote the safety measure of a car based on its position, speed, etc. If $y =$ danger and $a =$ safe, then the loss $L$(danger, safe) would be significantly high. This is because if the car is not within the safe region and the monitoring system wrongly estimates it, there could be a serious damage. However, if $y =$ safe but $a =$ danger, then the loss $L$(safe, danger) would have less impact. This is because even though the estimation is wrong, the car is still within the safe region, hence, it does not impact much.
}

The well-known loss functions such as 0-1 loss, quadratic loss, and logarithmic loss cannot address safety issues based on situational awareness within critical systems. The loss function $L$ in \eqref{loss} is more general than the existing loss functions.
By designing this unified loss function $L$ that incorporates the knowledge of the surrounding situation along with age, we can effectively capture and tackle safety-critical issues.

\vspace{-10pt}
\subsection{Information-theoretic Metric for Situational Awareness} \label{information_theory_view}

The performance of the safety-critical system for sensor $n$ at time slot $t$ is defined by the expected loss for a given state $(\Delta_n (t)=\delta$, $X_{n, t-\Delta_n (t)} = x)$ which can be expressed as the following penalty function:
\begin{align} \label{loss}
q_n (\delta, x) = 
\mathbb{E} [& L (Y_{n,t}, \phi_n (\Delta_n (t), X_{n, t-\Delta_n (t)})|\nonumber\\ & \Delta_n (t)=\delta, X_{n, t-\Delta_n (t)} =x],
\end{align} 
where $\phi_n (\cdot, \cdot)$ is any function that maps from $\mathbb N \times \mathcal X$ to $\mathcal A$.
Now, consider the following optimization problem:
\begin{align} \label{loss_1}
\min_{a \in \mathcal{A}} \mathbb{E} [L (Y_{n,t}, a)| \Delta_n (t)=\delta, X_{n, t-\Delta_n (t)} =x].
\end{align}
Let $a = \phi_n ^{*} (\Delta_n (t), X_{n, t- \Delta_n (t)})$ be the optimal estimator that solves the optimal estimation problem in \eqref{loss_1}. By substituting this optimal estimator into \eqref{loss}, we get
\begin{align} \label{loss_2}
q_n (\delta, x) = \mathbb{E} [& L (Y_{n,t}, \phi_n ^{*} (\Delta_n (t), X_{n, t-\Delta_n (t)})| \nonumber\\ & \Delta_n (t)=\delta, X_{n, t-\Delta_n (t)} =x],
\end{align}
which is a lower bound of $\mathbb{E} [L (Y_{n,t}, a)| \Delta_n (t)=\delta, X_{n, t-\Delta_n (t)} =x]$ for any $a \in \mathcal A$.
This penalty function in \eqref{loss_2} is closely related to the concept of generalized entropy \cite{dawid1998coherent, farnia2016minimax} or specifically, the $L$-entropy \cite{Shisher2022} of a random variable $Y_{n,t}$ given by 
\begin{align}
H_L (Y_{n,t}) = \min _{a \in \mathcal{A}} \mathbb{E} [L (Y_{n,t}, a)].
\end{align}
Furthermore, $L$-conditional entropy of $Y_{n,t}$ given $\Delta_n (t)=\delta$  and $X_{n, t-\Delta_n (t)} = x$ can be defined as  \cite{dawid1998coherent, farnia2016minimax, Shisher2022}
\begin{align} \label{L_cond_en}
& H_L (Y_{n,t} | \Delta_n (t)=\delta, X_{n, t-\Delta_n (t)} =x) \nonumber\\
= & \min _{a \in \mathcal{A}} \mathbb{E} [L (Y_{n, t}, a)| \Delta_n (t)=\delta, X_{n, t-\Delta_n (t)} =x], \nonumber\\
=& \mathbb{E} [L (\!Y_{n, t}, \phi_n ^{*}\! (\!\Delta_n (t), X_{n, t- \Delta_n (t)}\!)\!)| \Delta_n (t)\!\!=\!\delta, X_{n, t-\Delta_n (t)}\!\! =\!x].
\end{align}
From \eqref{loss_2} and \eqref{L_cond_en}, it is evident that
\begin{align} \label{equality}
q_n (\delta, x) = H_L (Y_{n, t} | \Delta_n (t)=\delta, X_{n, t-\Delta_n (t)} =x).
\end{align}
For the optimal estimator $\phi_n ^{*} (\cdot, \cdot)$, 
$q_n (\delta, x)$ is indeed $L$-conditional entropy which is an information-theoretic lower bound of $q_n (\delta, x)$. It represents the fundamental performance limit that characterizes the performance degradation due to the lack of the knowledge of the situation. The proposed metrics in prior works, i.e, AoII, VoI, AoS, QAoI cannot explain this information theoretic bound. In addition, for any general estimator output $a \in \mathcal A$, $q_n (\delta, x)$ can be represented as the $L$-conditional cross-entropy. Due to space limitation, the details are relegated to our future submission.

\subsection{Non-monotonic Information Aging} \label{non_monotonic}

Our analysis reveals that $q_n (\delta, x)$ can be a non-monotonic function of the age, particularly when the knowledge of the surrounding situation is taken into consideration which is illustrated in Figure \ref{figure1}(b). To do this experiment, we consider a safety-critical system where $N$ robots are moving in a gridworld with $5$ rows and $12$ columns, demonstrated in Figure \ref{figure1}(a). The observed state $X_{n, t}$ of robot $n$ is represented by two variables: the position $S_{n, t}$ of robot $n$ at time $t$ and its moving direction $a_{n, t}$ at time $t$ and $Y_{n, t}=$ \emph{\{safe, cautious, dangerous\}} denotes the safety level. In the gridworld in Figure \ref{figure1}(a), the states $X_{n, t}$ in row 1,2, and 3 are \emph{safe}, row 4 states are \emph{cautious}, and row 5 states are \emph{dangerous}. Row 3 is close to the boundary region between \emph{safe} and \emph{cautious}. Let $(x, y)$ denotes the position of a robot where $x$ is the row and $y$ is the column. The available moving directions $a_{n, t}$ for row 2, 3, and 4 are \emph{up, down, left}, and \emph{right}. For row 1, the \emph{up} is not available and for row 5, the \emph{down} is not available. If robot $n$ is in the leftmost position, then $a_{n,t}=$ left means it will stay in the same position, similar criteria is applied for the rightmost position. The probability of moving from one row to the adjacent row is $0.05$ (\emph{up} or \emph{down}) and the probability of staying in the same row is $0.95$ (\emph{left} or \emph{right}). The losses considered in Figure \ref{figure1}(b) are: $L$\emph{(cautious, safe)} $= 50, L$\emph{(safe, cautious)} $= 10, L$\emph{(dangerous, safe)} $=200, L$\emph{(safe, dangerous)} $=10, L$\emph{(dangerous, cautious)} $=50$, $L$\emph{(cautious, dangerous)}$=20$, and $L$\emph{(dangerous, dangerous)} $=L$\emph{(cautious, cautious)} $=L$\emph{(safe, safe)} $=0$. We consider optimal estimator of \eqref{loss_1} in this experiment.

From Figure \ref{figure1}(b), we observe that when a robot is in a \emph{safe} region and far from the \emph{safe} and \emph{cautious} boundary which is represented by the curve for given $X_{n,t} = (1,3)$, \emph{right}, the penalty is initially close to zero for small AoI values and increases gradually with increasing age. This phenomenon tells us that we do not need to update frequently when a robot is far from the boundary region. However, if the robot moves closer to the boundary between \emph{safe} and \emph{cautious} that is represented by the curve for given $X_{n,t} = (2,3)$, \emph{down}, the penalty increases very quickly because of the uncertainty of its position in the subsequent time slots. With the increase in age, this curve approaches to its stationary distribution. In similar way, the other curves can be explained. This penalty curves are not necessarily monotonic with age. 
Hence, only considering the non-decreasing functions of the age is not sufficient for performance analysis of safety-critical systems. The proposed metrics in prior works, i.e, AoII, VoI, AoS, QAoI cannot explain this non-monotonicity with age.  
\subsection{Scheduling Policy and Problem Formulation}
Let the scheduling policy is denoted by $\pi = (\pi_n)_{n=1}^{N}$ where $\pi_n = (\mu_n (0), \mu_n (1), \ldots)$ determines whether an observation is requested from sensor $n$ at every time slot $t \in \mathbb{N}$. Let $\Pi$ denotes the set of all causal scheduling policies in which every decision is made by using the current and history information available at the receiver. Because our system consists of $M$ channels, $\sum_{n=1}^{N} \mu_n (t) \leq M$ is required to hold for all $t$.%

We aim to find an optimal scheduling policy that minimizes the time-average sum of expected penalty of the $N$ sources over an infinite time-horizon $T$, which is formulated as 
\begin{align}
\mathsf{q}_{\text{opt}} = & \inf_{\pi \in \Pi} \limsup_{T \to \infty}\!\! \sum_{n=1}^{N}  \!\mathbb{E}_{\pi} \!\!\bigg[\!\frac{1}{T} \!\!\sum_{t=0}^{T-1} q_n (\Delta_n (t), X_{n, t-\Delta_n (t)}) \bigg], \label{problem} \\
& ~\text{s.t.} \sum_{n=1}^{N} \mu_n (t) \leq M, \mu_n (t) \in \{0,1\}, t = 0,1, \ldots,\label{constraint}
\end{align}
where $q_n (\Delta_n (t), X_{n, t-\Delta_n (t)})$ is the penalty incurred by source $n$ at time $t$ which is defined in \eqref{loss}, and $\mathsf{q}_{\text{opt}}$ is the optimum value of \eqref{problem}. 

\section{Penalty-minimization: An Information-theoretic View}

In Section \ref{information_theory_view}, we demonstrate that the penalty function $q_n (\delta, x)$ can be represented as $L$-conditional entropy. Leveraging this insight, we obtain that for optimal estimators, always sending updates benefits the system by reducing its average penalty of the system. To prove this result, we present the following useful lemma which illustrates that more information reduces the $L$-conditional entropy.

\begin{lemma} \label{lemma1}
For random variables $X, Y,$ and $Z$, it holds that $H_L (Y|Z=z) \geq H_L (Y|X, Z=z)$, where
\begin{align}
H_L (Y|Z=z) &= \min_{a \in \mathcal{A}} \mathbb{E} [L(Y,a)|Z=z], \label{entropy_eq_1}\\
H_L (Y|X, Z\!=\!z) \!\!&= \!\!\sum_{x \in \mathcal{X}} P(X\!=\!x|Z\!=\!z) H_L (Y|X\!=\!x, Z\!=\!z). \label{entropy_eq_2}
\end{align}
\end{lemma}


Then we have the following theorem.


\begin{theorem} \label{theorem1}
If the packet transmission times are one-time slot, then for optimal estimators it is always better to keep the channels busy.
\end{theorem}

Due to space limitation, the proofs of Lemma \ref{lemma1} and Theorem \ref{theorem1} are relegated to our future submission.

Because problem \eqref{problem}-\eqref{constraint} has a channel resource constraint, all of the sensors cannot submit their updates at every time slot when $N > M$. Therefore, we have to design an efficient scheduling policy that minimizes the time-average sum of the expected penalty of the $N$ sources ensuring that constraint \eqref{constraint} is satisfied. We provide the details in the next section. 
\section{Restless Multi-armed Bandit Formulation}

Problem \eqref{problem}-\eqref{constraint} is an RMAB problem where each source $n$ is an arm and $(\Delta_n (t), X_{n, t- \Delta_n (t)})$ is the state of each arm $n$. To find an optimal solution to the RMAB problem is PSPACE hard \cite{PSPACE_book}. A Whittle index policy is known to be asymptotically optimal for many RMAB problems \cite{weber1990index}. However, it needs to satisfy a complicated condition called indexability. Due to the complicated nature of the state transitions and non-monotonic age-penalty functions along with erasure channels, it is difficult to establish indexability for our problem. Therefore, in this work, we provide a low-complexity algorithm that does not need to satisfy indexability. Next, we demonstrate that the developed policy is asymptotically optimal.

\subsection{Relaxation and Lagrangian Decomposition}

Following the standard relaxation and Lagrangian decomposition procedure for RMAB \cite{whittle_restless}, the original problem in \eqref{problem}-\eqref{constraint} is relaxed as 
\begin{align} 
\mathsf{q}_{\text{opt}}\!\! =\!\! & \inf_{\pi \in \Pi} \!\!\limsup_{T \to \infty} \! \sum_{n=1}^{N}\!  \mathbb{E}_{\pi}\! \bigg[\frac{1}{T} \!\sum_{t=0}^{T-1} q_n (\Delta_n (t), X_{n, t-\Delta_n (t)}) \bigg], \label{relaxed_problem}  \\
& ~\text{s.t.} \limsup_{T \to \infty}  \sum_{n=1}^{N} \mathbb{E}_{\pi} \bigg[\frac{1}{T} \sum_{t=0}^{T-1} \mu_n (t) \bigg] \leq M, \label{relaxed_cons_2}
\end{align}
where the relaxed constraint \eqref{relaxed_cons_2} only needs to be satisfied on average, whereas \eqref{constraint} is required to hold at any time $t$.
To solve the relaxed problem \eqref{relaxed_problem}-\eqref{relaxed_cons_2}, we take a dual cost $\lambda \geq 0$ (also knows as Lagrange multiplier) for the relaxed constraint. The dual problem is given by
\begin{align} \label{dual_problem}
\sup_{\lambda \geq 0} \bar q (\lambda), 
\end{align}
where 
\begin{align} \label{decomposed}
\bar q (\lambda) = \inf_{\pi \in \Pi} \limsup_{T \to \infty} \sum_{n=1}^{N} \mathbb{E}_{\pi} \bigg[& \frac{1}{T} \sum_{t=0}^{T-1}  q_n (\Delta_n (t), X_{n, t-\Delta_n (t)}) \nonumber\\
& + \lambda \big( \mu_n (t) - M\big) \bigg].
\end{align}
The term $\frac{1}{T} \sum_{t=0}^{T-1} \sum_{n=1}^{N} \lambda M$ in \eqref{decomposed} does not depend on policy $\pi$ and hence can be removed. For a given $\lambda$, problem \eqref{decomposed} can be decomposed into $N$ separated sub-problems and each sub-problem associated with source $n$ is formulated as
\begin{align} \label{per_arm_problem}
\bar q_n (\lambda) = \inf_{\pi_n \in \Pi_n} \limsup_{T \to \infty} \mathbb{E}_{\pi_n} \bigg[\frac{1}{T} \sum_{t=0}^{T-1} & q_n (\Delta_n (t), X_{n, t-\Delta_n (t)}) \nonumber\\
& + \lambda \mu_n (t) \bigg],
\end{align}
where $\bar q_n (\lambda)$ is the optimum value of \eqref{per_arm_problem}, $\pi_n= (\mu_n (1), 
\mu_n (2), \ldots)$ denotes a sub-scheduling policy for source $n$, and $\Pi_n$ is the set of all causal sub-scheduling policies of source $n$. 

\begin{algorithm}[t] 
\caption{Net-gain Maximization Policy} \label{algo1}
\begin{algorithmic}[1]
\State At time $t=0$:
\State Input $\lambda^{*}$ which is the optimal solution to \eqref{dual_problem}.
\State Input $\alpha_{n, \lambda^{*}} (\delta, x)$ in \eqref{net_gain} for every source $n$. 
\State For all time $t = 0, 1, \ldots$,
\State Update $(\Delta_n (t), X_{n, t-\Delta_n (t)})$ for all source $n$.
\State Update current ``gain" $\alpha_{n, \lambda^{*}} (\Delta_n (t), X_{n, t-\Delta_n (t)})$ for all source $n$.
\State Choose at most $M$ sensors with highest positive ``gain". 
\end{algorithmic}
\end{algorithm}

\section{Optimal Policy via Dynamic Programming}

Given transmission cost $\lambda$, the per-arm problem \eqref{per_arm_problem} is an average-cost infinite horizon MDP with state $(\Delta_n (t), X_{n, t-\Delta_n (t)})$. We solve \eqref{per_arm_problem} by using dynamic programming \cite{bertsekas2011dynamic}. 
The Bellman optimality equation for the MDP in \eqref{per_arm_problem} is 
\begin{align}
h_{n, \lambda} (\delta, x) = \min_{\mu \in \{0,1\}} Q_{n, \lambda} (\delta, x, \mu),
\end{align}
where $h_{n, \lambda} (\delta, x)$ is the relative-value function of the average-cost MDP and $Q_{n, \lambda} (\delta, x, \mu)$ is the relative action-value function defined as
\begin{align} \label{action_function}
& Q_{n, \lambda} (\delta, x, \mu) = \nonumber\\
& \begin{cases}
q_n (\delta, x) - \bar q_n (\lambda) + h_{n, \lambda} (\delta+1, x) , &\!\!\!\!\!\!\!\!\! \text{if} {\thinspace} \mu =0, \\
q_n (\delta, x) - \bar q_n (\lambda) + (1-p_n) h_{n, \lambda} (\delta+1, x) \\
+ p_n  \mathbb{E} [h_{n, \lambda} (1, X_{n, t-1}) | X_{n, t-\delta} = x]+ \lambda,  
& \!\!\!\!\!\!\!\!\! \text{otherwise}.
\end{cases}
\end{align}
The relative-value function $h_{n, \lambda} (\delta, x)$ can be computed by using relative value iteration algorithm for average-cost MDP \cite{bertsekas2011dynamic}.
Following \cite{chen2022index, shisher2023learning}, define the ``gain" $\alpha_{n, \lambda} (\delta, x)$ for choosing the action $\mu_{n, \lambda} (t)$ as
\begin{align} \label{net_gain}
\alpha_{n, \lambda} (\delta, x) = Q_{n, \lambda} (\delta, x, 0) - Q_{n, \lambda} (\delta, x, 1).
\end{align}
Substituting \eqref{action_function} into \eqref{net_gain}, we get
\begin{align} \label{gain_1}
& \alpha_{n, \lambda} (\delta, x) = \nonumber\\
& p_{n}\! \bigg(\!\!h_{n, \lambda}\! (\delta\!+\!1, x) \!-\!
 \mathbb{E} [h_{n, \lambda}\! (1, X_{n, t-1}) | X_{n, t-\delta} = x]\! \bigg)\!-\!  \lambda.
\end{align}
By utilizing the ``gain" $\alpha_{n, \lambda} (\delta, x)$ in \eqref{gain_1}, we obtain the optimal decision to the relaxed problem \eqref{relaxed_problem}-\eqref{relaxed_cons_2} at time $t$ for every sensor $n$ as
\begin{align}
\mu_n (t) = \argmax_{\mu \in \{0,1\}} \alpha_{n, \lambda (t)} (\Delta_n (t), X_{n, t-\Delta_n (t)}, \mu),
\end{align}
where the dual cost is iteratively updated using the dual subgradient ascent method with step size $\beta > 0$ \cite{nedic2008subgradient}:
\begin{align} \label{lambda_update}
\lambda (t+1) = \max \bigg\{\lambda (t) + \beta/t \bigg(\sum_{n=1}^{N} \mu_n (t) - M\bigg), 0\bigg\}.
\end{align}
Let $\lambda^ *$ be the optimal dual cost to problem \eqref{dual_problem} to which $\lambda (t)$ converges. 
We provide a low-complexity algorithm for solving problem \eqref{problem}-\eqref{constraint} in Algorithm \ref{algo1}. We utilize the ``gain” defined in \eqref{net_gain} as the priority measurement for choosing action $\mu$. Algorithm \ref{algo1} takes optimal dual cost $\lambda^{*}$ and the precomputed gain $\alpha_{n, \lambda^{*}} (\delta, x)$ associated with $\lambda^{*}$ as input. Then, for all $t \geq 0$, the state $(\Delta_n (t), X_{n, t-\Delta_n (t)})$ and the associated ``gain" $\alpha_{n, \lambda^ {*}} (\Delta_n (t), X_{n, t-\Delta_n (t)})$ are updated. Finally, Algorithm \ref{algo1} maximizes the ``Net-gain” (total gain of all sensors) of the system at time $t$. This is done by selecting at most $M$ sensors having the highest positive “gain” at time $t$. 
The ``Net-gain Maximization Policy” in Algorithm \ref{algo1} does not need to satisfy the indexability condition. 


\section{Asymptotic Optimality}
In this section, we demonstrate that the "Net-gain Maximization Policy" in Algorithm \ref{algo1} is asymptotically optimal in the same asymptotic regime as the Whittle index policy \cite{whittle_restless}. In this scenario, all $N$ arms are generalized to $N$ classes, and the
number of arms in each class and the number of channels $M$ are scaled by $\gamma$, while maintaining a constant ratio between them.

Let $Z_n ^{\gamma} (\pi^{*}, \{\delta, x\}, \mu, t)$ be a fluid-scaling process with parameter $\gamma$ that represents the expected number of class-$n$ arms at state $(\delta, x)$ that takes action $\mu$ at time slot $t$ under policy $\pi^{*}$. Consider the following expected long-term average cost
\begin{align}
V_{\pi^{*}} ^{\gamma} = & \limsup_{T \to \infty} \frac{1}{T} \mathbb{E}_{\pi^{*}}\! \bigg[\sum_{t=0}^{T-1} \sum_{n=1}^{N} \!\!\!\sum_{\substack{(\Delta_n (t), \\X_{n, t-\Delta_n (t)})}} \!\!\!\!\!\! q_n (\Delta_n (t), X_{n, t-\Delta_n (t)}) \nonumber\\
& \frac{Z_n ^{\gamma} (\pi^{*}, \{\Delta_n (t), X_{n, t-\Delta_n (t)}\}, \mu, t)}{\gamma}\bigg].
\end{align}
The policy $\pi^{*}$ will be asymptotically optimal if $V_{\pi^{*}} ^{\gamma} \leq V_{\pi} ^{\gamma}$ for all $\pi \in \Pi$. In this sequel, we introduce the following \emph{global attractor} \cite{verloop2016asymptotically}.

\begin{definition} \label{def2}
\textbf{Global attractor.} An equilibrium point ${Z_n ^{\gamma, *}}/{\gamma}$ under policy $\pi^{*}$ is a global attractor for the process ${Z_n ^{\gamma} (\pi^{*}; t)}/{\gamma}$, if, for any initial point $ {Z_n ^{\gamma} (\pi^{*}; 0)}/{\gamma}$, the process ${Z_n ^{\gamma} (\pi^{*}; t)}/{\gamma}$ converges to ${Z_n ^{\gamma, *}}/{\gamma}$.
\end{definition}


\begin{theorem} \label{optimality}
Under Definition \ref{def2}, the policy $\pi^{*}$ is asymptotically optimal. Therefore, $\lim_{\gamma \to \infty} V_{\pi^{*}} ^{\gamma} = V_{\pi^{\text{opt}}} ^{\gamma}$, where $\pi^{\text{opt}}$ is the optimal policy for the original RMAB problem \eqref{problem}-\eqref{constraint}. 
\end{theorem}

Due to space limitation, the proof of Theorem \ref{optimality} is relegated to our future submission.
\section{Numerical Results} \label{simulation}

\begin{figure}
\vspace{-0.3cm}
\centering
\includegraphics[width=6.7cm]{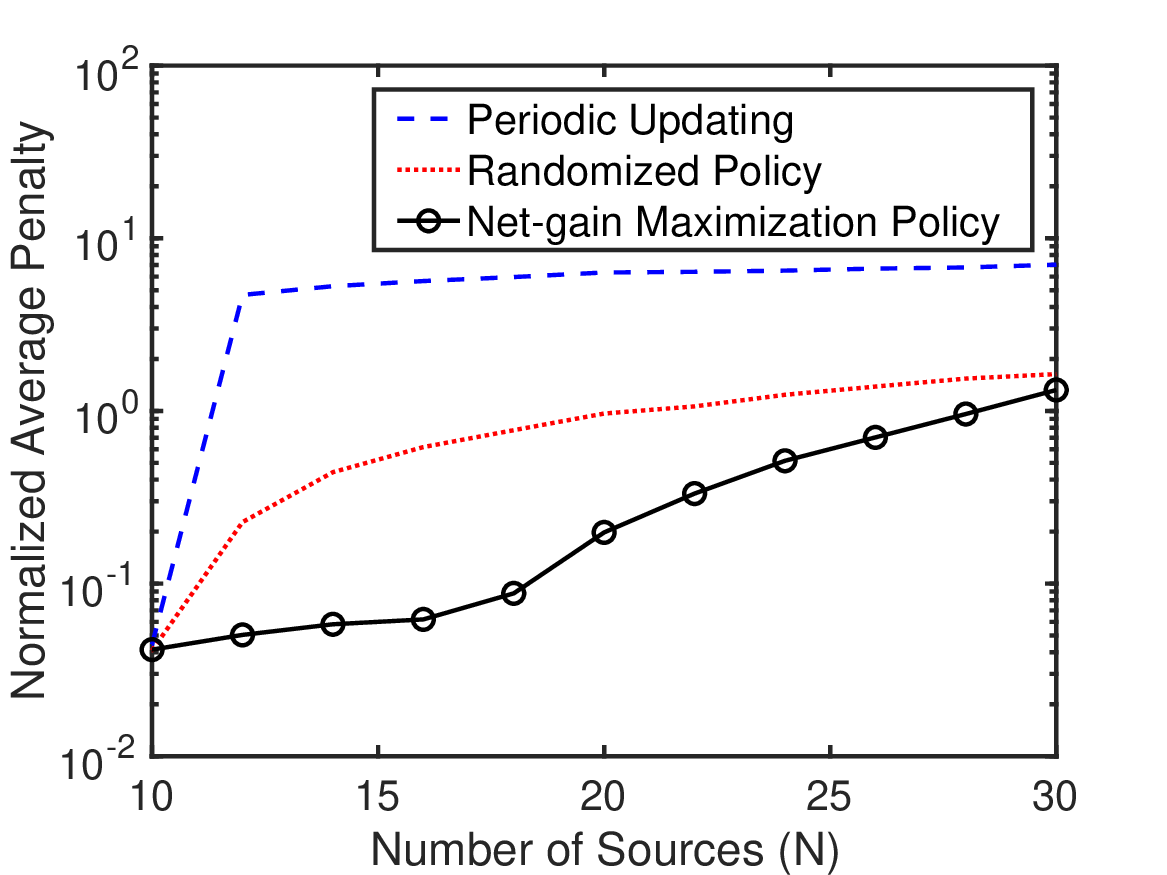}  
\vspace*{2.0mm} 
\caption{Normalized average penalty vs Number of sources ($N$) where Number of channels are $M=10$ with success probability $0.95$.}\vspace{-0.0cm}
\label{figure2}
\end{figure}  

In this section, we evaluate the performance of the following policies:

\begin{itemize}

\item Periodic Updating: The sensors generate updates at every time slot and store in a FIFO queue. Whenever a channel is available, an update from the queue is sent.

\item Randomized Policy: If $M$ channel resources are available, this policy randomly selects at most $M$ sensors.

\item Net-gain Maximization Policy: See Algorithm \ref{algo1}.
 
\end{itemize}

We consider the same experimental setup of Figure \ref{figure1} where $5$ robots follow a deterministic policy (they follow a fixed path). 
The cost associated with these $5$ robots is zero at every time slot because given an initial state, the position of these robots can be uniquely determined by following the deterministic policy. The goal of the other $N-5$ robots is to move and scan the environment (e.g., Mars Rovers \cite{muirhead2004mars}) and send updates when requested. We do not consider any termination state for these robots, the goal is to keep scanning for infinite-time horizon. 
Our system consists of $M=10$ erasure channels and the success probability is $0.95$.

The performance comparison of the three policies mentioned above is provided in Figure \ref{figure2}. The normalized average penalty in Figure \ref{figure2} is obtained by dividing time-average cost by the number of robots. From the figure, until $N \leq M$, all of the three policies show the same performance. Whenever $N > M$, periodic updating starts getting worse because the queue length is getting higher. In our simulation, we have used a buffer size of 20 for periodic updating. Moreover, the randomized policy randomly selects at most $5$ sensors for sending updates, whereas the net-gain maximization policy makes the decision in a smarter way by considering the AoI and the state of the surrounding situation. The performance gain of the net-gain maximization policy is up to 100 times compared to periodic updating and up to 10 times compared to the randomized policy.

\section{conclusion}
We address the importance of situational awareness in safety-critical systems. The general loss function $L$ have practical importance and appropriate design of $L$ can address many safety-critical issues. In future we will study systems where multiple sensors can arrive and leave the system at any time. Another interesting direction is to consider a finite time horizon problem where there is a termination state while encountering a danger. 
\bibliographystyle{IEEEtran}
\bibliography{ref,ref1,ref_2,sueh}

\begin{thebibliography}{10}
\providecommand{\url}[1]{#1}
\csname url@samestyle\endcsname
\providecommand{\newblock}{\relax}
\providecommand{\bibinfo}[2]{#2}
\providecommand{\BIBentrySTDinterwordspacing}{\spaceskip=0pt\relax}
\providecommand{\BIBentryALTinterwordstretchfactor}{4}
\providecommand{\BIBentryALTinterwordspacing}{\spaceskip=\fontdimen2\font plus
\BIBentryALTinterwordstretchfactor\fontdimen3\font minus
  \fontdimen4\font\relax}
\providecommand{\BIBforeignlanguage}[2]{{%
\expandafter\ifx\csname l@#1\endcsname\relax
\typeout{** WARNING: IEEEtran.bst: No hyphenation pattern has been}%
\typeout{** loaded for the language `#1'. Using the pattern for}%
\typeout{** the default language instead.}%
\else
\language=\csname l@#1\endcsname
\fi
#2}}
\providecommand{\BIBdecl}{\relax}
\BIBdecl

\bibitem{grau2017industrial}
A.~Grau, M.~Indri, L.~L. Bello, and T.~Sauter, ``Industrial robotics in factory
  automation: From the early stage to the internet of things,'' in \emph{IEEE
  IECON}, 2017, pp. 6159--6164.

\bibitem{abdulmalek2022iot}
S.~Abdulmalek, A.~Nasir, W.~A. Jabbar, M.~A. Almuhaya, A.~K. Bairagi, M.~A.-M.
  Khan, and S.-H. Kee, ``{I}o{T}-based healthcare-monitoring system towards
  improving quality of life: A review,'' in \emph{Healthcare}, vol.~10, no.~10,
  2022, p. 1993.

\bibitem{seenivasan2015disaster}
M.~Seenivasan, M.~Arularasu, K.~Senthilkumar, and R.~Thirumalai, ``Disaster
  prevention and control management in automation: a key role in safety
  engineering,'' \emph{Procedia Earth and Planetary Science}, vol.~11, pp.
  557--565, 2015.

\bibitem{li2020waiting}
F.~Li, Y.~Sang, Z.~Liu, B.~Li, H.~Wu, and B.~Ji, ``Waiting but not aging:
  Optimizing information freshness under the pull model,'' \emph{IEEE/ACM
  Trans. Netw.}, vol.~29, no.~1, pp. 465--478, 2020.

\bibitem{KaulYatesGruteser-Infocom2012}
S.~Kaul, R.~D. Yates, and M.~Gruteser, ``Real-time status: How often should one
  update?'' in \emph{IEEE INFOCOM}, 2012.

\bibitem{dawid1998coherent}
A.~P. Dawid, ``Coherent measures of discrepancy, uncertainty and dependence,
  with applications to bayesian predictive experimental design,''
  \emph{Department of Statistical Science, University College London}, vol.
  139, 1998.

\bibitem{farnia2016minimax}
F.~Farnia and D.~Tse, ``A minimax approach to supervised learning,''
  \emph{Advances in Neural Information Processing Systems}, vol.~29, 2016.

\bibitem{Shisher2022}
M.~K.~C. Shisher and Y.~Sun, ``How does data freshness affect real-time
  supervised learning?'' in \emph{ACM MobiHoc}, 2022, pp. 31--40.

\bibitem{Ali_AoII}
A.~Maatouk, S.~Kriouile, M.~Assaad, and A.~Ephremides, ``The age of incorrect
  information: A new performance metric for status updates,'' \emph{IEEE/ACM
  Trans. Netw.}, vol.~28, p. 2215–2228, oct 2020.

\bibitem{zhong2018two}
J.~Zhong, R.~D. Yates, and E.~Soljanin, ``Two freshness metrics for local cache
  refresh,'' in \emph{IEEE ISIT}, 2018, pp. 1924--1928.

\bibitem{zheng2020urgency}
X.~Zheng, S.~Zhou, and Z.~Niu, ``Urgency of information for context-aware
  timely status updates in remote control systems,'' \emph{IEEE Trans. Wirel.
  Commun.}, vol.~19, no.~11, pp. 7237--7250, 2020.

\bibitem{yates2021age}
R.~D. Yates, ``The age of gossip in networks,'' in \emph{IEEE ISIT}, 2021, pp.
  2984--2989.

\bibitem{holm2021freshness}
J.~Holm, A.~E. Kal{\o}r, F.~Chiariotti, B.~Soret, S.~K. Jensen, T.~B. Pedersen,
  and P.~Popovski, ``Freshness on demand: Optimizing age of information for the
  query process,'' in \emph{IEEE ICC}, 2021, pp. 1--6.

\bibitem{VoI_Kosta}
A.~Kosta, N.~Pappas, A.~Ephremides, and V.~Angelakis, ``Age and value of
  information: Non-linear age case,'' in \emph{IEEE ISIT}, 2017, pp. 326--330.

\bibitem{Chen_2022}
G.~Chen, S.~C. Liew, and Y.~Shao, ``Uncertainty-of-information scheduling: A
  restless multi-armed bandit framework,'' \emph{{IEEE} Trans. Inf. Theory},
  2022.

\bibitem{shisher2021age}
M.~K.~C. Shisher, H.~Qin, L.~Yang, F.~Yan, and Y.~Sun, ``The age of correlated
  features in supervised learning based forecasting,'' in \emph{IEEE INFOCOM
  Workshops}, 2021, pp. 1--8.

\bibitem{bertsekas2011dynamic}
D.~P. Bertsekas \emph{et~al.}, ``Dynamic programming and optimal control 3rd
  edition, volume ii,'' \emph{Belmont, MA: Athena Scientific}, vol.~1, 2011.

\bibitem{chen2022index}
G.~Chen and S.~C. Liew, ``An index policy for minimizing the
  uncertainty-of-information of {Markov} sources,'' \emph{arXiv preprint
  arXiv:2212.02752}, 2022.

\bibitem{SunNonlinear2019}
Y.~Sun and B.~Cyr, ``Sampling for data freshness optimization: Non-linear age
  functions,'' \emph{J. Commun. Netw.}, vol.~21, pp. 204--219, 2019.

\bibitem{Sun_TIT_2017}
Y.~Sun, E.~Uysal-Biyikoglu, R.~D. Yates, C.~E. Koksal, and N.~B. Shroff,
  ``Update or wait: How to keep your data fresh,'' \emph{IEEE Trans. Inf.
  Theory}, vol.~63, no.~11, pp. 7492--7508, 2017.

\bibitem{Bedewy_2021}
A.~M. Bedewy, Y.~Sun, S.~Kompella, and N.~B. Shroff, ``Optimal sampling and
  scheduling for timely status updates in multi-source networks,'' \emph{IEEE
  Trans. Inf. Theory}, vol.~67, no.~6, pp. 4019--4034, 2021.

\bibitem{Ornee2021}
T.~Z. Ornee and Y.~Sun, ``Sampling and remote estimation for the
  ornstein-uhlenbeck process through queues: Age of information and beyond,''
  \emph{IEEE/ACM Trans. Netw.}, vol.~29, no.~5, p. 1962–1975, oct 2021.

\bibitem{Wiener_TIT}
Y.~{Sun}, Y.~{Polyanskiy}, and E.~{Uysal}, ``Sampling of the {Wiener} process
  for remote estimation over a channel with random delay,'' \emph{IEEE Trans.
  Inf. Theory}, vol.~66, no.~2, pp. 1118--1135, 2020.

\bibitem{jsac_survey}
R.~D. Yates, Y.~Sun, D.~R. Brown, S.~K. Kaul, E.~Modiano, and S.~Ulukus, ``Age
  of information: An introduction and survey,'' \emph{IEEE J. Sel. Areas
  Commun.}, vol.~39, no.~5, pp. 1183--1210, 2021.

\bibitem{pappas2021goal}
N.~Pappas and M.~Kountouris, ``Goal-oriented communication for real-time
  tracking in autonomous systems,'' in \emph{IEEE ICAS}, 2021, pp. 1--5.

\bibitem{wang2022framework}
Z.~Wang, M.-A. Badiu, and J.~P. Coon, ``A framework for characterizing the
  value of information in hidden {Markov} models,'' \emph{IEEE Trans. Inf.
  Theory}, vol.~68, no.~8, pp. 5203--5216, 2022.

\bibitem{soleymani2016optimal}
T.~Soleymani, S.~Hirche, and J.~S. Baras, ``Optimal self-driven sampling for
  estimation based on value of information,'' in \emph{IEEE WODES}, 2016, pp.
  183--188.

\bibitem{shisher2023learning}
M.~K.~C. Shisher, B.~Ji, I.~Hou, Y.~Sun \emph{et~al.}, ``Learning and
  communications co-design for remote inference systems: Feature length
  selection and transmission scheduling,'' \emph{arXiv preprint
  arXiv:2308.10094}, 2023.

\bibitem{kadota2018scheduling}
I.~Kadota, A.~Sinha, E.~Uysal-Biyikoglu, R.~Singh, and E.~Modiano, ``Scheduling
  policies for minimizing age of information in broadcast wireless networks,''
  \emph{IEEE/ACM Trans. Netw.}, vol.~26, no.~6, pp. 2637--2650, 2018.

\bibitem{hsu2018age}
Y.-P. Hsu, ``Age of information: Whittle index for scheduling stochastic
  arrivals,'' in \emph{IEEE ISIT}, 2018, pp. 2634--2638.

\bibitem{ornee2023whittle}
T.~Z. Ornee and Y.~Sun, ``A {Whittle} index policy for the remote estimation of
  multiple continuous {Gauss-Markov} processes over parallel channels,''
  accepted by ACM MobiHoc 2023.

\bibitem{xiong2022index}
G.~Xiong, X.~Qin, B.~Li, R.~Singh, and J.~Li, ``Index-aware reinforcement
  learning for adaptive video streaming at the wireless edge,'' in \emph{ACM
  MobiHoc}, 2022, pp. 81--90.

\bibitem{zou2021minimizing}
Y.~Zou, K.~T. Kim, X.~Lin, and M.~Chiang, ``Minimizing age-of-information in
  heterogeneous multi-channel systems: A new partial-index approach,'' in
  \emph{ACM MobiHoc}, 2021, pp. 11--20.

\bibitem{chen2021scheduling}
Y.~Chen and A.~Ephremides, ``Scheduling to minimize age of incorrect
  information with imperfect channel state information,'' \emph{Entropy},
  vol.~23, no.~12, p. 1572, 2021.

\bibitem{yun2023remote}
J.~Yun, A.~Eryilmaz, J.~Moon, and C.~Joo, ``Remote estimation for dynamic {IoT}
  sources under sublinear communication costs,'' \emph{IEEE/ACM Trans. Netw.},
  2023.

\bibitem{PSPACE_book}
C.~Papadimitriou and J.~Tsitsiklis, ``The complexity of optimal queueing
  network control,'' in \emph{IEEE CCC}, 1994, pp. 318--322.

\bibitem{weber1990index}
R.~R. Weber and G.~Weiss, ``On an index policy for restless bandits,''
  \emph{Journal of applied probability}, vol.~27, no.~3, pp. 637--648, 1990.

\bibitem{whittle_restless}
P.~Whittle, ``Restless bandits: activity allocation in a changing world,''
  \emph{Journal of Applied Probability}, vol. 25A, pp. 287--298, 1988.

\bibitem{nedic2008subgradient}
A.~Nedic and A.~Ozdaglar, ``Subgradient methods in network resource allocation:
  Rate analysis,'' in \emph{IEEE CISS}, 2008, pp. 1189--1194.

\bibitem{verloop2016asymptotically}
I.~M. Verloop, ``Asymptotically optimal priority policies for indexable and
  nonindexable restless bandits,'' 2016.

\bibitem{muirhead2004mars}
B.~K. Muirhead, ``Mars rovers, past and future,'' in \emph{IEEE aerospace
  conference}, vol.~1, 2004.

\end{thebibliography}

\end{document}